\documentclass[twocolumn,noshowpacs,preprintnumbers,amsmath,amssymb]{revtex4-1}
\usepackage{graphicx}
\usepackage{dcolumn}
\usepackage{bm}

\begin{document}

\title{Nonlocal long-range synchronization of planar Josephson junction arrays}

\author{S. Yu. Grebenchuk$^{1,2}$}
\author{R. Cattaneo$^1$}
\author{T. Golod$^1$}
\author{V. M. Krasnov$^{1,2,3}$}
\email{Vladimir.Krasnov@fysik.su.se}

\affiliation{$^1$ Department of Physics, Stockholm University,
AlbaNova University Center, SE-10691 Stockholm, Sweden;}

\affiliation{$^2$ Moscow Institute of Physics and Technology,
141700 Dolgoprudny, Russia;}

\affiliation{$^3$ Institute for Physics of Microstructures RAS, Nizhny Novgorod 603950, Russia}

\begin{abstract}
We study arrays of 
planar Nb Josephson junctions with contacts to intermediate electrodes, which allow measurements of individual junctions and, thus, provide an insight into intricate array dynamics. We observe a robust phase-locking of arrays, despite a significant inter-junction separation. Several unusual phenomena are reported, such as a bi-stable critical current with re-entrant superconductivity upon switching of nearby junctions; and ``incorrect" Shapiro steps, occurring at mixing frequencies between the external RF radiation and the internal Josephson frequency in nearby junctions. Our results reveal 
a surprisingly strong and long-range inter-junction interaction. It is attributed to nonlocality of planar junction electrodynamics, caused by the long-range spreading of stray electromagnetic fields. The nonlocality greatly enhances the high-frequency interjunction coupling and enables large-scale synchronization. 
Therefore, we conclude that planar geometry is advantageous for realization of coherent Josephson electronics.

\end{abstract}

\maketitle

\section{Introduction}

Coherent operation of several phase-locked Josephson junctions (JJs) can significantly improve performance of Josephson cryoelectronic devices \cite{Jain_1984}. For example, it can enhance power efficiency of Josephson oscillators 
\cite{Jain_1984,Han_1994,Barbara_1999,Kleiner_2000,Filatrella_2006,Ozyuzer_2007,Benseman_2013,Welp_2013,Borodianskyi_2017,Galin_2018,Cattaneo_2021},
reduce fluctuations and improve stability \cite{Jain_1984,Mros_1998}, and increase (multiply) the readout voltage 
\cite{Klushin_1995,Ravindran_1996,Cybart_2019,Golod_2019}. 
The coherence is vital because incoherent operation of multi-junction devices can instead deteriorate the performance \cite{Krasnov_2002}. Phase-locking 
is facilitated by inter-junction coupling. It can be direct, or indirect. 
A direct coupling can be mediated by non-equilibrium effects associated with quasiparticle injection \cite{Jain_1984}, or by common magnetic 
\cite{Sakai_1993,Kleiner_1994} and electric 
\cite{Koyama_1996,Shukrinov_2007} fields shared by JJs. The corresponding decay lengths in superconductors are short (nano-scale). The quasiparticle relaxation length is in the range 10-100 nm (unless very close to $T_c$) \cite{Jain_1984}, the screening length of magnetic field is determined by the London penetration depth, $\lambda_L\sim 100$ nm, and of the electric field - by the Debye 
length $< 1$ nm. Therefore, the profound direct coupling usually occurs only in stacked JJs with very thin common electrodes \cite{Ustinov_1993,Nevirkovets_1994}, being particularly strong 
in atom-scale intrinsic JJs in high-$T_c$ cuprates \cite{Koyama_1996,Mros_1998,Shukrinov_2007,Kleiner_2000,Ozyuzer_2007,Benseman_2013,Welp_2013,Borodianskyi_2017,Cattaneo_2021,Sakai_1993,Kleiner_1994,Koyama_1996,Shukrinov_2007}. Indirect coupling occurs via individual interaction with a common external resonator \cite{Barbara_1999,Almaas_2002,Galin_2018,Galin_2020}. In this case the separation between JJs is not critical and can be much larger than the screening lengths \cite{Jain_1984,Barbara_1999,Galin_2018,Galin_2020}. 

Synchronization is essentially a dynamic process. It implies locking of time-dependent Josephson phase differences, $\varphi$, 
accompanied by voltage-locking of JJ's \cite{Jain_1984,Kohlstedt_1995,Darula_1995,Goldobin_1996}. 
Current-locking, i.e. identical critical currents, $I_c$, for switching from the static, 
$V=0$, into the resistive 
state, is also commonly observed \cite{Ustinov_1993,Nevirkovets_1994,Goldobin_1996}. However, we want to emphasize that pure static phase-locking is not possible. Indeed, the static phase is determined by the dc-Josephson relation, $I=I_c\sin(\varphi)$. Since $I_c$ of JJs 
are not absolutely identical, the phases of serially biased JJs can not be equal \cite{Jain_1984}. Consequently, current locking may either be caused by a dynamic interaction: the first switched JJ induces ac-currents that drag other JJs into the resistive state; or it may be of thermal origin: the switched JJ heats the device reduces $I_c$ of other JJs and causes their switching. 

Tracking of individual JJs is neccessary for understanding of the synchronization process. This requires access to intermediate electrodes in the array \cite{Nevirkovets_1994,Kohlstedt_1995,Darula_1995,Goldobin_1996}. The latter is technically difficult for vertical stacks of conventional overlap-type JJs \cite{Nevirkovets_1994,Kohlstedt_1995}, but easy for planar JJs \cite{Darula_1995,Golod_2019}. The two-dimensional geometry of planar JJs alters their physical properties. 
Most importantly, planar JJs have nonlocal electrodynamics \cite{Mintz_2001,Boris_2013} due to the presence of stray electric and magnetic fields \cite{Golod_2019B}. As we will show below, this 
strongly affects mutual coupling between planar JJs.    

Here we study arrays of planar Nb JJs with access to intermediate electrodes. This facilitates analysis of individual JJs and small junction groups. A robust synchronization of arrays, leading to straightforward multiplication of the read-out $I_c R_n$ voltage, is observed. 
We report a strong dynamic cross-talking between JJs: JJs mutually affect voltages and critical currents of each other. Large inter-junction currents lead to unusual behavior with bi-stable $I_c$ and re-entrant superconductivity; and to formation of ``incorrect" Shapiro steps, corresponding to mixing frequencies between external RF radiation and internal Josephson oscillations. 
Our results reveal a surprisingly strong coupling between JJs despite a significant ($\gg\lambda_L$) inter-junction separation. 
It is explained by the nonlocal electrodynamics of planar JJs, facilitating direct long-range coupling. We conclude that the planar geometry is beneficial for creation of large-scale phase-coherent Josephson electronics. 

\begin{figure*}[t]
    \centering
    \includegraphics[width=0.99\textwidth]{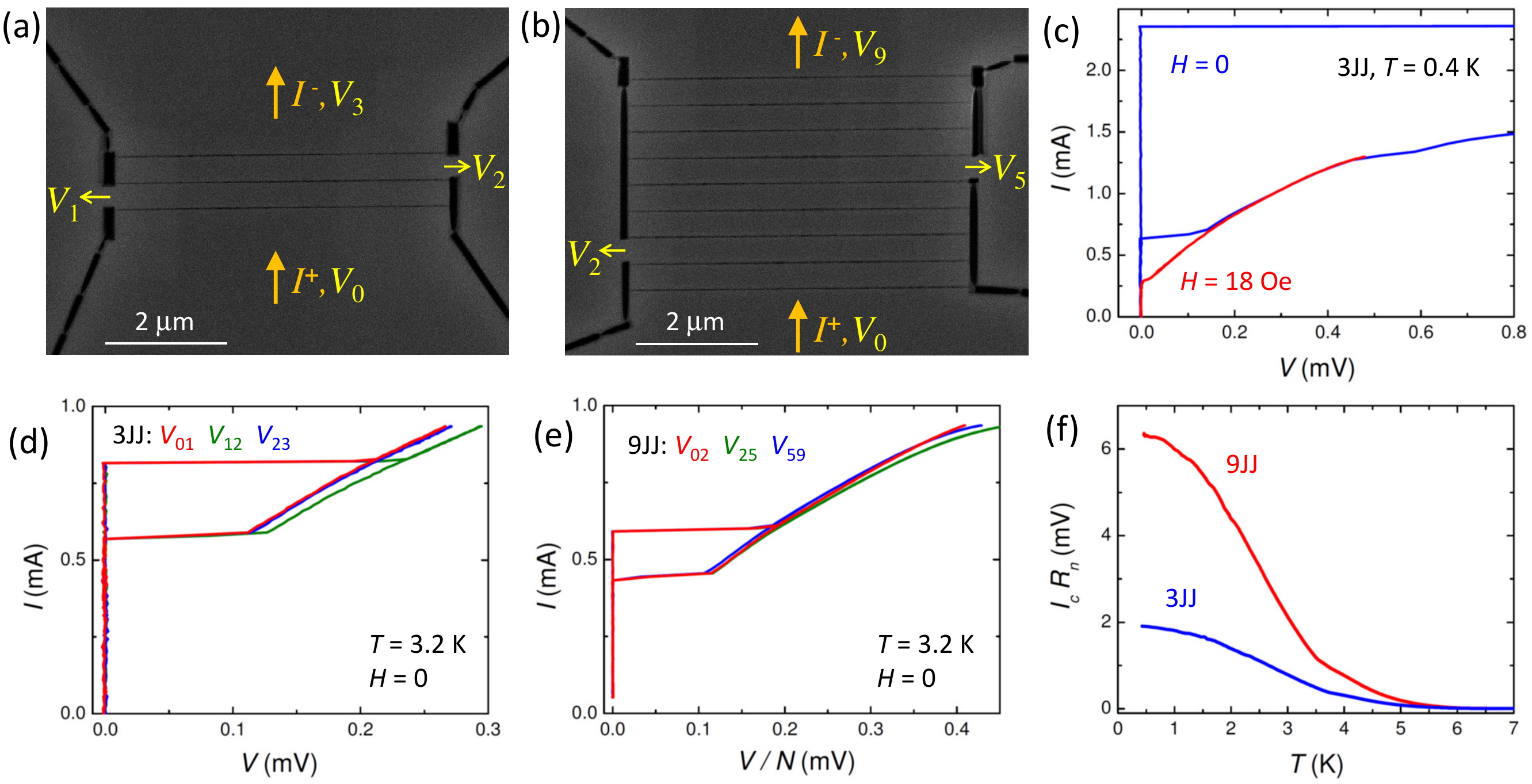}
    \caption{(Color online). (a) and (b) SEM images of the two studied arrays with (a) three and (b) nine planar Nb JJs. Each array has two additional contacts to middle electrodes, marked by arrows at the left and right sides of the array. (c) The $I$-$V$ characteristics at $T=0.4$ K for one of the JJs in the 3JJ array at zero (blue) and high (red) fields. (d) and (e) The $I$-$V$s of (d) individual JJs in the 3JJ array and (e) three  sections of the 9JJ array, normalized by the number of JJs, $N$, in each section. (f) Temperature dependencies of the total $I_c R_n$ for both arrays. Multiplication of the readout voltage is clearly seen. }
    \label{fig:fig1}
\end{figure*}

\section{Results}

Figures \ref{fig:fig1} (a) and (b) represent scanning electron microscope (SEM) images of studied arrays. Variable thickness-type planar JJs are made by Ga$^+$ focused ion beam etching of a 70 nm thick Nb film. JJs can be seen as horizontal lines in the SEM images. 
The 3JJ array (a) contains three, and the 9JJ array (b) nine JJs. The JJ width is, $W \simeq 5~\mu$m, and the inter-junction separation, $L\simeq 0.5~\mu$m.  We numerate JJs from bottom to top: JJ 1, 2, 3, ... 
The fabrication procedure is similar to that described earlier \cite{Krasnov_2005,Golod_2019,Golod_2019B}, but here we use a single Nb film instead of a bilayers. This allows significant enhancement of the characteristic $I_c R_n$ voltage ($R_n$ is the quasiparticle resistance), which reaches $\sim 0.7$ mV at low temperatures. This can be seen from the blue current-voltage ($I$-$V$) characteristics in Fig. \ref{fig:fig1} (c) obtained at low $T$ and zero field 
for one of the JJs in the 3JJ array. JJs have overdamped characteristics with an excess current, typical for constriction-type weak links \cite{Jain_1984,Octavio_1988}. 
This can be seen from the red $I$-$V$ in Fig. \ref{fig:fig1} (c), obtained at high field with maximally suppressed $I_c$. 
At high bias the $I$-$V$s become nonlinear, presumably due to a resistive (flux-flow type) contribution from Nb electrodes \cite{Kapran_2021}. To avoid ambiguity, we define $R_n$ from the linear low-bias part of the $I$-$V$. Measurements are performed in a closed-cycle cryostat. Magnetic field is applied perpendicular to Nb electrodes. 

\begin{figure*}[t]
    \centering
    \includegraphics[width=0.99\textwidth]{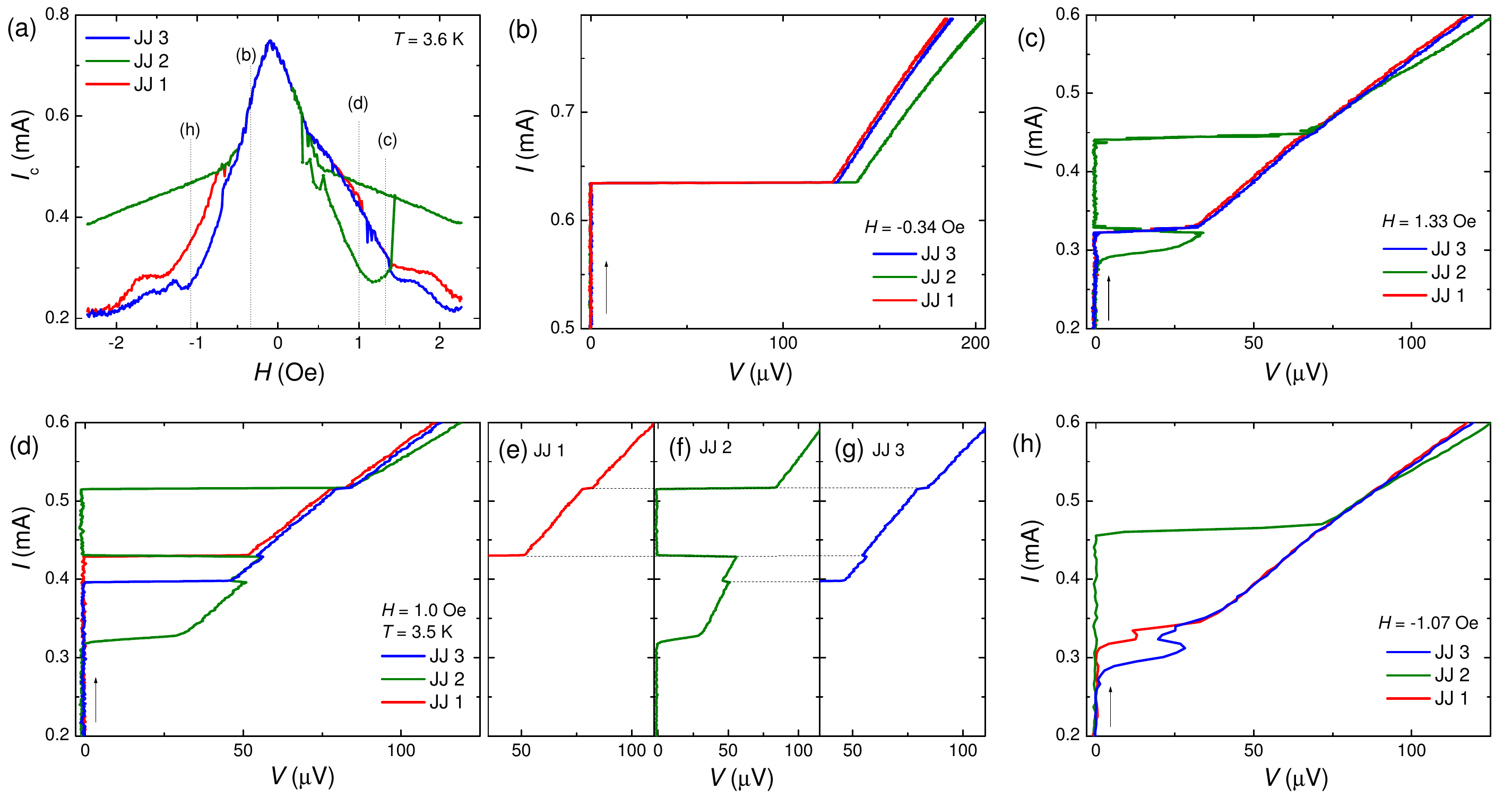}
    \caption{(Color online). (a) Magnetic field dependencies of critical (switching) currents of JJs in the 3JJ array. A clear current locking of all JJs is seen at -0.5 Oe $\lesssim H \lesssim $ 0.5 Oe. Bi-stable region of $I_c$ occurs at 0.5 Oe $\lesssim H \lesssim $ 1.5 Oe. Panels (b-h) show the $I$-$V$s of JJs at four magnetic fields, marked by vertical dotted lines in (a). Panel (b) shows the current-locked state at small fields. (c) Corresponds to both current and voltage locking in JJs 1 and 3 and a bi-stable state in JJ 2. Panels (d-g) show another meta-stable state with re-entrant superconductivity in JJ 2. Insets (e-g) demonstrate a strong dynamic cross-talking between JJs in the resistive state. Panel (h) illustrates the occurrence of a strong mutual interaction between outmost JJs 1 and 3, which leads to a robust voltage-locking in the dynamic state. }
    \label{fig:fig2}
\end{figure*}
 
Planar geometry facilitates a simple access to intermediate electrodes. Studied arrays have two additional contacts marked by left and right arrows in Figs. \ref{fig:fig1} (a) and (b). JJs are biased in series from the bottom, $I^+$, to the top, $I^-$, electrode. 
Voltages are measured simultaneously between all contacts. For the 3JJ array 
we can measure independently each JJ: $V_{JJ 1}=V_0-V_1$, $V_{JJ 2}=V_1-V_2$, $V_{JJ 3}=V_2-V_3$, as shown in Fig. \ref{fig:fig1} (d). 
For the 9JJ array we can measure the bottom section containing JJs 1,2: $V_{02}=V_0-V_2$, the middle section with three JJs 3-5: $V_{25}=V_2-V_5$, and the top section with four JJs 6-9: $V_{59}=V_5-V_9$. 
Fig. \ref{fig:fig1} (e) shows corresponding $I$-$V$s, normalized by the number of JJs, $N$, in each section. From Figs. \ref{fig:fig1} (d) and (e) it is seen that JJs have similar characteristics, demonstrating good reproducibility of the fabrication procedure. 
Fig. \ref{fig:fig1} (f) shows temperature dependencies of the total $I_c R_n$ at $H=0$ for both arrays. 
The multiplication effect of the array readout voltage is clearly seen with the $I_c R_n$ reaching 2 mV for the 3JJ and more than 6 mV for the 9JJ array.  

\subsection{Magnetic field (de)tuning}

From Figs. \ref{fig:fig1} (d) and (f) it is seen that $I_c$s of JJs in each array are identical. To understand whether such current-locking is due to synchronization, or just a coincidence, we detune JJs by applying magnetic field. Unlike conventional overlap JJs, the effective flux quantization area, $A$, of planar JJs is determined by the length and width of electrodes \cite{Mintz_2001,Boris_2013,Golod_2019}. For inner JJs with similar narrow, $L<W$, electrodes $A=LW/2$. For outer JJs with dissimilar electrodes $A=[W^2/1.8 + WL/2]/2$ \cite{Golod_2019}. As a result, the flux quanization field, $\Delta H=\Phi_0/A$, of inner JJs is approximately 6 times larger than for outmost JJs, i.e., inner JJs are six times less sensitive to magnetic field. This allows detuning by magnetic field. Figure \ref{fig:fig2} (a) shows $I_c(H)$ modulation for each JJ in the 3JJ array. Current-locking of all JJs is clearly seen in the range $-0.5$ Oe $\lesssim H \lesssim 0.5$ Oe, followed by a partial locking of two JJs and unlocked states at higher fields.  

\begin{figure*}[t]
    \centering
    \includegraphics[width=0.99\textwidth]{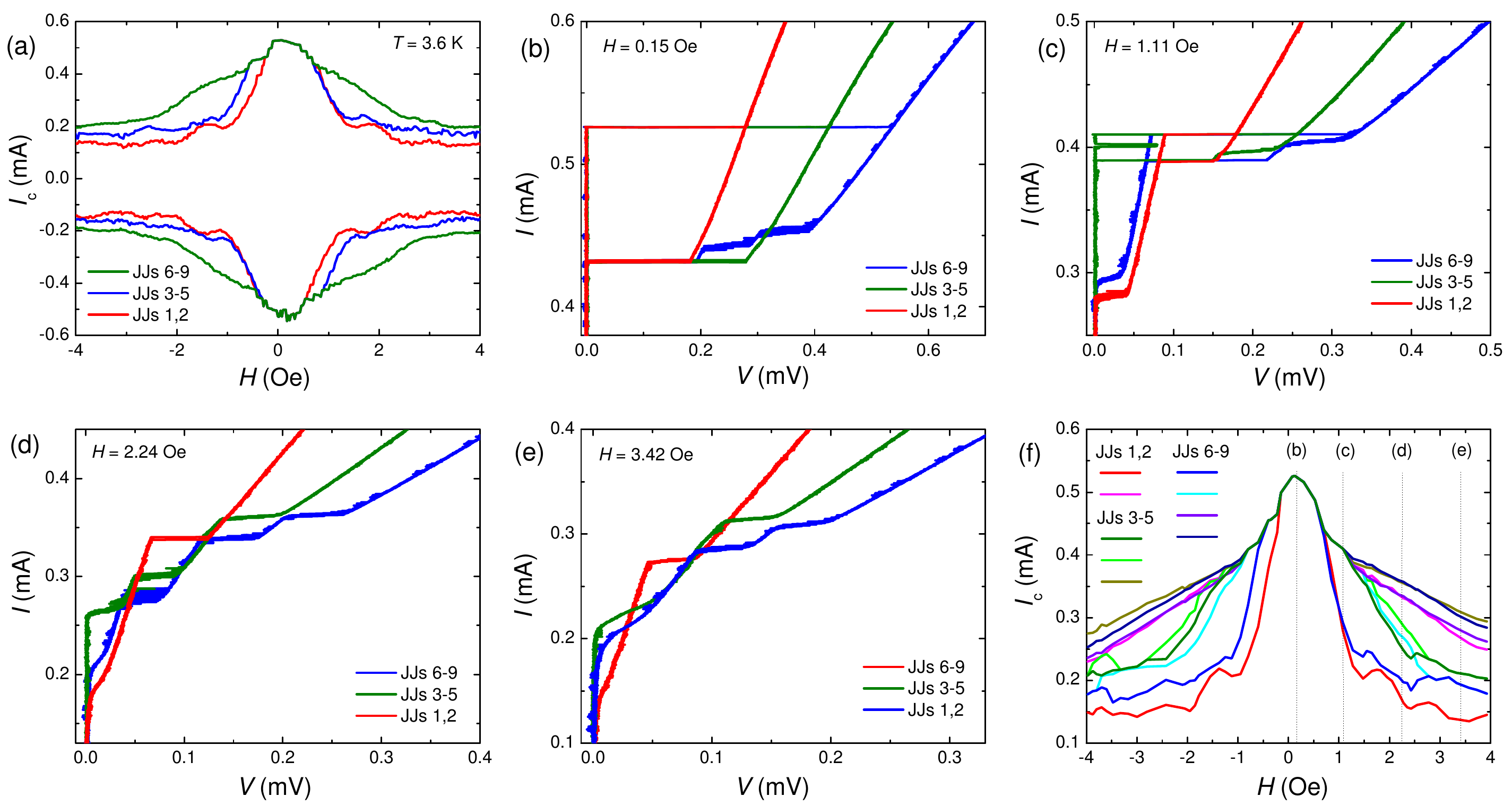}
    \caption{(Color online). (a). $I_c(H)$ modulation of three sections in the 9JJ array. (b-e) The $I$-$V$s of array sections at four magnetic fields, indicated in (f). (f) $I_c(H)$ modulation of all JJs in the 9JJ array, deduced from the $I$-$V$s. }
    \label{fig:fig3}
\end{figure*}

\subsection{Bi-stable re-entrant critical current}

As we emphasized above, phase-locking is essentially a dynamic phenomenon. Therefore, current-locking can not be purely static. For getting insight into the synchronization process, $I$-$V$s of individual JJs should be analyzed. The Supplementary video-1 \cite{Supplem} shows magnetic field variation of the $I$-$V$s in the 3JJ array. It reveals an intricate dynamic behavior of the array. 
Figs. \ref{fig:fig2} (b-f) show snapshots at four magnetic fields indicated by vertical lines in Fig. \ref{fig:fig2} (a). Fig. \ref{fig:fig2} (b) corresponds to the current-locked state at small field. Despite that, JJs do not voltage-lock in the resistive state. Synchronization is hindered by a slightly higher $R_n$ of JJ 2. 

Fig. \ref{fig:fig2} (c) shows a partially-locked case at a higher field. Here outer JJs 1, 3 are firmly phase-locked (both current- and voltage- locked), but the inner JJ 2 shows a remarkable bi-stable behavior with re-entrant superconductivity. When JJs 1 and 3 are in the superconducting state it has a low $I_c\simeq 0.28$ mA. However, when JJs 1,3 switch in the resistive state, JJ 2 acquires a significantly higher $I_c^*\simeq 0.45$ mA and returns back to the superconducting state. When JJ 2 switches into the resistive state at $I>I_c^*$, all JJs get voltage-locked up to $I\simeq 0.5$ mA, at which JJ 2 returns to its higher $R_n$. 

Fig. \ref{fig:fig2} (d) represents the case without current-locking at $V=0$. Here JJs have different $I_c$s and switch one-by-one. This allows a clear observation of dynamic cross-talking between JJs. Panels (e-g) show the individual $I$-$V$s. First, JJ 2 switches at $I\simeq 0.32$ mA. When JJ 3 follows at $I\simeq 0.4$ mA, JJs 2 and 3 get firmly voltage-locked, which is accompanied by the abrupt reduction of voltage in JJ 2. 
However, as JJ 1 switches at $I\simeq 0.43$ mA, JJ 2 returns to the superconducting, $V=0$, state and JJs 1 and 3 get (nearly) voltage-locked, which is accompanied by theabrupt reduction of voltage in JJ 3. Finally, at $I\simeq 0.52$ mA, JJ 2 returns to the resistive state, followed by jumps in voltages of JJs 1 and 3. Thus, switching of one JJ 
affects both $V$ and $I_c$ of all other JJs, not only nearest neighbors. The corresponding voltage changes are abrupt, which is the signature of phase-locking processes \cite{Jain_1984}. Fig. \ref{fig:fig2} (h) demonstrates that even outmost JJs 1 and 3, which do not share common electrodes, exhibit a strong dynamic cross-talking. Here JJ 3 switches first at $I\simeq 0.28$ mA. When JJ 1 follows at $I\simeq 0.31$ mA, voltages of both JJs 1 and 3 get adjusted although they are not direct neighbors. After that JJs 1 and 3 become robustly voltage-locked in a broad current range.    

In Figure \ref{fig:fig3} and the Supplementary video-2 \cite{Supplem} we show similar data for the 9JJ array. Panel (a) shows $I_c(H)$ for the three sections of the array. The general behavior is similar to the 3JJ array with a clear current-locking at the central lobe, as shown in panel (b). The range of locking is similar to the 3JJ array despite a three-times larger array size. This indicates a long-range nature of interaction between planar JJs. Beyond the current-locking range, the $I$-$V$s contain multiple steps, corresponding to switching of individual JJs, as can be seen from Figs. \ref{fig:fig3} (c-e). Thus we can measure $I_c(H)$ modulation of all individual JJs, as shown in Fig. \ref{fig:fig3} (f). 
It reveals various regions of both global and partial current-locking of JJs in the array. This array also shows some bi-stable behavior, as can be seen from Fig. \ref{fig:fig3} (c). 

\begin{figure*}[t]
    \centering
    \includegraphics[width=0.99\textwidth]{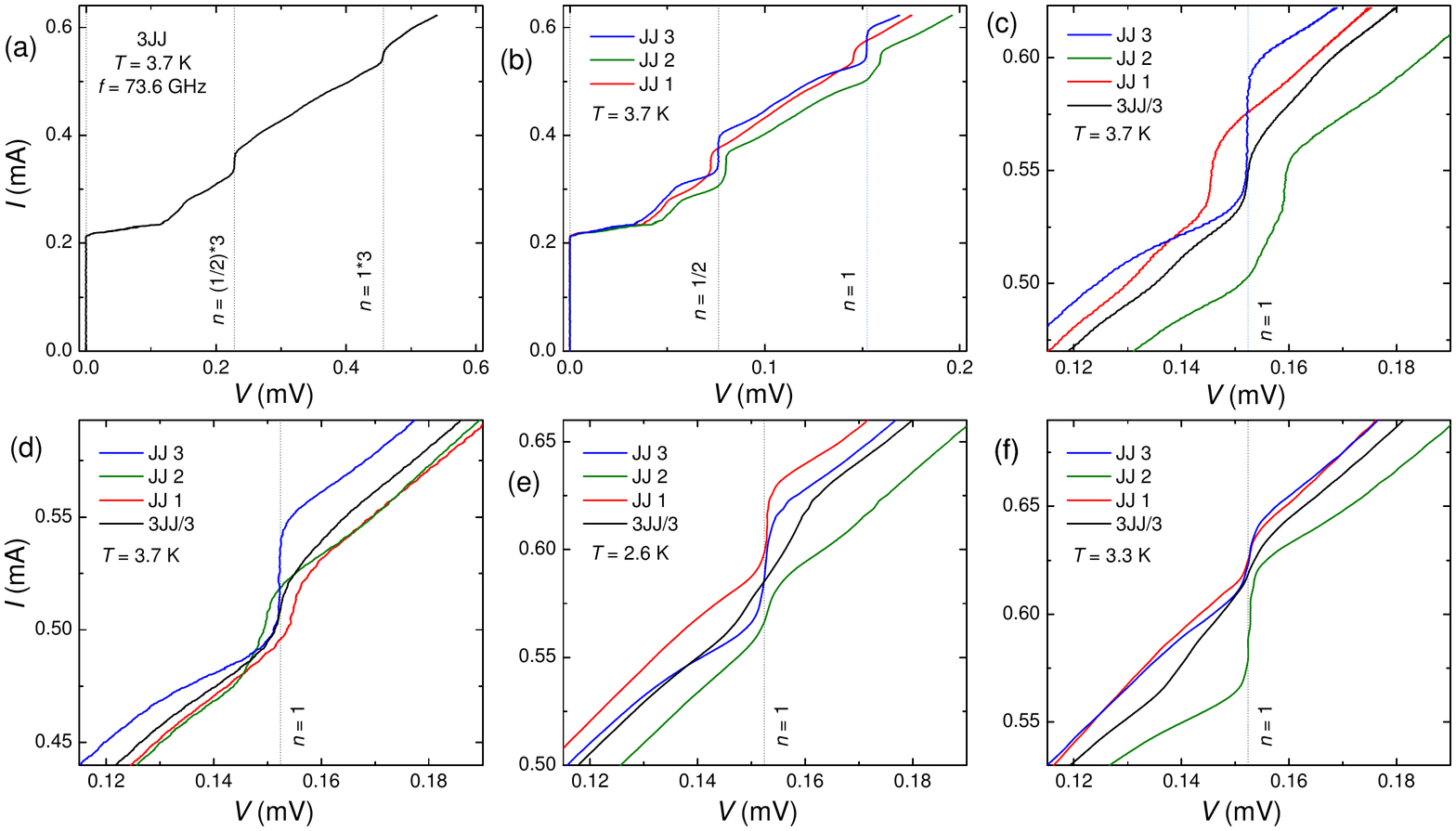}
    \caption{(Color online). (a) The $I$-$V$s of the 3JJ array irradiated by $f_{RF}=73.6$ GHz. Giant Shapiro steps are seen for $n=1$ and $n=1/2$. (b) $I$-$V$s of individual JJs. (c) Zoom-in into the $n=1$ step region. It is seen that only JJ 3 exhibits the correct Shapiro step, while in JJs 1 and 2 steps appear at ``incorrect" voltages, corresponding to mixing frequencies with Josephson oscillations. This indicates existence of large induced high-frequency currents from neighbor junctions. Panel (d) represents the case similar to (c) in a slightly higher RF power. It is seen that the voltage difference between JJ 1 and 3 is smaller than in (c). Panels (e) and (f) represent different unsynchronized states at lower $T$. Here there are no giant steps in the overall $I$-$V$ (black lines), but each JJ has an ordinary ``correct" Shapiro step, albeit with largely unequal amplitudes. }
    \label{fig:fig4}
\end{figure*}

\subsection{RF response with ``incorrect" Shapiro steps}

The dynamic cross-talking reported above provides the key evidence for a strong mutual coupling: a JJ in the dynamic state affects all other JJs in the array. This implies that Josephson oscillations from one JJ induce significant alternating currents in other JJs. Technically, this is similar to the case when a junction is subjected to an external RF radiation. It is known that RF irradiation leads to appearance of Shapiro steps in the $I$-$V$ \cite{Barone}. 
The corresponding additional dc-current (of either sign) is caused by rectification of the RF-current due to nonlinearity of junction response \cite{Jain_1984}. The observed dynamic cross-talking between JJs implies that ac-currents, induced by Josephson oscillations in one JJ, penetrate in other JJs and get partly rectified. This allows adjustment of the dc $I$-$V$s and enables voltage-locking of dissimilar JJs \cite{Jain_1984}. 

To verify this scenario, we studied RF response of the arrays. 
In Fig. \ref{fig:fig4} (a-c) we show the $I$-$V$s of the 3JJ array irradiated by $f_{RF}\simeq 73.6$ GHz at $T=3.7$ K. It is expected that giant Shapiro steps at $V_n=nN(hf/2e)$, where $N$ is the number of JJs and $n$ is the Shapiro step number, should appear in the synchronized state 
\cite{Klushin_1995,Ravindran_1996}. Indeed, doted lines in Figs. \ref{fig:fig4} (a) indicate well developed integer, $n=1$, and a half-integer, $n=1/2$, giant steps. Sub-harmonic steps are caused by a non-sinusoidal current-phase relationship in constriction JJs \cite{Kulik_1978}. It may seem that JJs are synchronized at the giant Shapiro steps. However, the $I$-$V$s of individual JJs, shown in Figs. \ref{fig:fig4} (b,c) reveal that this is not the case. Remarkably, although Shapiro-like steps are well developed in each JJ, they have different voltages. This is more obvious from Fig. \ref{fig:fig4} (c), which shows a closeup of the $n=1$ step. It is seen that only JJ 3 exhibits a regular Shapiro step at $V_1=hf_{RF}/2e$, while in JJ 1 and 2 steps appear at ``incorrect" voltages $V_1-\delta V$ and $V_1+\delta V$ with $\delta V\simeq 6.9~\mu$V. Fig. \ref{fig:fig4} (d) represents a similar state at a slightly higher RF power. Here $\delta V\simeq 2.6~\mu$V is reduced, 
indicating that the RF power tries to synchronize JJs but doesn't quit succeed due to variation of junction properties. Finally, in Figs. \ref{fig:fig4} (e) and (f) we show two unsynchronized states at lower $T$. Here there are no giant steps in the overall $I$-$V$s (black lines), however, each JJ has a conventional Shapiro step at the correct voltage. 

Note that in all shown cases, Figs. \ref{fig:fig4} (c-f), the step amplitudes vary significantly between JJs. E.g., in Figs \ref{fig:fig4} (c) and (d) the step is large in JJ 3 and small in JJs 1 and 2; in Figs \ref{fig:fig4} (e) it is large  in JJs 3 and 1 and small in JJ 2; in (f) it is large in JJ 2 and tiny in JJs 1 and 3. Such variation is surprising because the separation between JJs is much smaller that the RF wavelength $\sim 4$ mm. Therefore, JJs experience the same incoming RF amplitude. Furthermore, it can not be said that some particular JJ is poorly responding to RF because large steps can be seen in all JJs. Therefore, the observed surprisingly large variation of Shapiro step amplitudes indicates that junction responses are influenced by nearby JJs as much as by the external RF power.

Appearance of ``incorrect" Shapiro steps, Figs. \ref{fig:fig4} (c) and (d), provides simultaneously the most spectacular (we are unaware of similar earlier reports) and clear evidence 
for existence of additional high-frequency currents induced by JJs themselves. The symmetrical arrangement of these steps in JJs 1 and 2 with respect to the correct Shapiro step in JJ 3 implies that the ``incorrect" steps involve mixing components with the frequency $\delta f = f_{3} - f_1 = f_2-f_{3}$, where $f_i$ is the Josephson frequency in JJ $i$, so that $f_3=f_{RF}$, $f_1= f_{RF}-\delta f$ and $f_2= f_{RF}+\delta f$. As a result, the giant steps in the overall $I$-$V$, see black lines in Figs. \ref{fig:fig4} (c) and (d), appear at a correct voltage. But, our observation demonstrates that the giant step itself does not prove array synchronization. 

For the 9JJ array RF-induced steps were also observed, but amplitudes of giant steps were much smaller. This indicates that the external RF power was not effective in synchronizing JJs. This was not because of insufficient RF power (it was high enough to completely quench critical currents), but presumably because of overwhelming frequency mixing effects. Appearance of Shapiro steps at a beat frequency in a two-junction case has been observed \cite{Jain_1984}. A possible reason for appearance of ``incorrect" steps in our arrays could be a so called ``mixing with gain", which can occur in overdamped JJs \cite{Jain_1984}. This phenomenon amplifies low-frequency mixing components $f\ll I_c R_n/2he$ due to the large differential resistance $dV/dI$ in the $I$-$V$ of overdamped JJs at $V \ll I_c R_n$. This may lead to a large amplitude of low-frequency beat components with $\delta f \simeq 3.3$ and 1.3 GHz in cases of Figs. \ref{fig:fig4} (c) and (d), respectively. 

\section{Discussion}

We have shown that the ability to measure $I$-$V$s of individual JJs is instrumental for understanding an intricate array dynamics in the phase-locking process. 
As can be seen from Figs. \ref{fig:fig2} (c) and (d), voltage-locking occurs despite a larger $R_n$ in JJ 2. As we emphasized above, this must involve ac-currents induced by neighbor JJs. Rectification of such currents shifts the $I$-$V$ by the mechanism similar to formation of Shapiro steps upon external RF-radiation. 
Therefore, to estimate the amplitude of induced ac-currents we compare the current shift, $\Delta I$, required for voltage-locking, with the current size of Shapiro steps. Since the rectified dc-current at a Shapiro step changes from negative to positive along the step, the maximal rectified current is half the total Shapiro step size. 

From comparison of Figs. \ref{fig:fig2} (a) and \ref{fig:fig4} (a) it follows that the RF radiation suppresses $I_c$ to $\lesssim 0.3~I_{c0}$ ($I_{c0}$ is the unperturbed value). 
This corresponds to RF- currents $I_{RF}\sim I_{c0}$ \cite{Barone}. On the other hand, from Fig. \ref{fig:fig2} (c) it can be seen that voltage-locking of JJ-2 requires a dc-shift $\Delta I \simeq 20~\mu$A at $V\simeq 75~\mu$V. A similar shift is seen in Fig. \ref{fig:fig2} (f) upon voltage-locking of JJs 2 and 3 
at $I\simeq 0.4$ mA. Such $\Delta I$ is about 50$\%$ of the half-size of the $n=1/2$ Shapiro step, which occurs at similar voltages, see Fig. \ref{fig:fig4} (a). Therefore, the neighbor-induced ac-current is $\sim 0.5 I_{c0}$. Since the amplitude of Josephson oscillations is $I_{c0}$, the ac-current is reduced only by half. This is quite surprising, taking into account that the separation between JJs (500 nm) is significantly larger than $\lambda_L$ and larger than the Pearl length \cite{Golod_2019B}. For comparison, a similar range of current-locking in stacked JJs could be observed only at JJ separation $\sim 10$ nm $\ll \lambda_L$ \cite{Ustinov_1993,Nevirkovets_1994,Kohlstedt_1995,Goldobin_1996}. Furthermore, the current-locking range for 3JJ and 9JJ arrays is similar, despite a factor three difference in size, see Figs. \ref{fig:fig2} (a) and \ref{fig:fig3} (a). Moreover, in Ref. \cite{Darula_1995} synchronization of planar high-$T_c$ 
JJs with $\sim 10~\mu$m separation was reported. 
All this indicates the involvement of a well known dynamic high-frequency electromagnetic coupling mechanism \cite{Jain_1984}, the only peculiarity of which for our planar JJ arrays is its long-range nature.

On the other hand, the reported re-entrant behavior of $I_c$ provides evidence for existence of another, static, coupling mechanism. From Figs. \ref{fig:fig2} (c) and (f) it is seen that $I_c$ in JJ 2 is almost doubled after switching of JJs 1,3. 
This 
can not be explained solely by induced ac-currents. Indeed, $I_c$ decreases with increasing $I_{RF}$ \cite{Barone}. Although oscillatory behavior appears at higher amplitudes $I_{RF}>I_{c0}$, the $I_c(I_{RF})$ maxima never exceed $I_{c0}$. This is in contrast to the reported re-entrant behavior with larger $I_c$ in the dynamic state. 
Therefore, the bi-stability must have another origin. A some-what similar hysteretic re-entrant behavior has been reported for superconductor/ferromagnet hybrid structures upon re-orientation of local magnetization in the ferromagnet \cite{Zdravkov_2010,Blamire_2014,Birge_2016,Kapran_2021}. This demonstrates that the change of $I_c$ can be due to a changed static magnetic background. In our case the most likely cause of that is entrance or exit of Josephson vortices (fluxons) 
\cite{Dremov_2019}, 
triggered by switching of neighbor JJs. The corresponding fluxon-induced metastability is well studied for stacked JJs \cite{Mros_1998}. Fluxon-related origin of re-entrance in our arrays is consistent with observation that it appears at $H\sim 1$ Oe, close to the field for penetration of the first fluxon in outmost JJs [the first minimum in red and blue $I_c(H)$ curves from Fig. \ref{fig:fig3} (f)].   

\begin{figure}[t]
    \centering
    \includegraphics[width=0.49\textwidth]{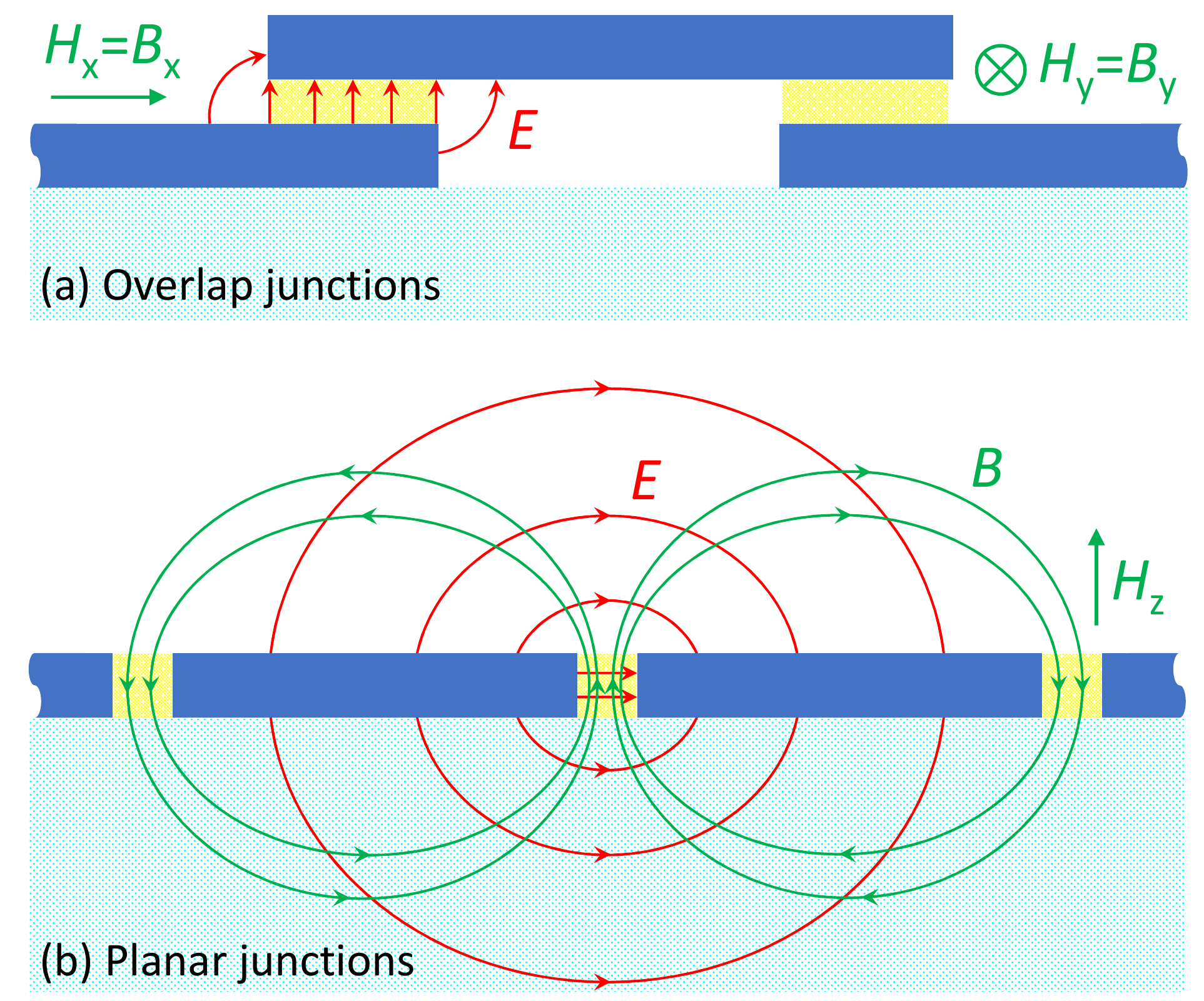}
    \caption{(Color online). A sketch of distribution of electric and magnetic fields in arrays of (a) conventional overlap and (b) planar JJs. The key difference is the absence/presence of long-range stray fields in (a)/(b), caused by (a) parallel-plate and (b) coplanar capacitor geometries of the two types of JJs. }
    \label{fig:fig5}
\end{figure}

Thus we identified two mechanisms of long-range coupling between planar JJs: (i) a dynamic coupling via induced ac-currents and related electromagnetic fields; and (ii) quasi-static coupling via magnetic fields of Josephson vortices. Both are direct coupling mechanisms (our arrays don't have any obvious resonator required for indirect coupling). The range of such direct coupling, $\sim \mu$m, in planar JJs is remarkably long compared to $\sim 10$ nm in conventional overlap JJs \cite{Ustinov_1993,Nevirkovets_1994,Kohlstedt_1995,Goldobin_1996}. 

To clarify the difference and explain the origin of such direct long-range coupling, in Figure \ref{fig:fig5} we sketched geometries of (a) overlap and (b) planar junction arrays.
Overlap JJs have a parallel-plate capacitor geometry. Electric field is concentrated in the capacitor and only tiny edge fields leave the junctions. This makes interaction via electric field weak and short range since the characteristic distance is determined by the separation between electrodes, which is in the range of 10 nm. Magnetic field is applied parallel to electrodes. Since the demagnetization factor in this case is close to zero, there are no stray magnetic fields. Consequently, there is no direct magnetic interaction, provided the inter-junction separation is much larger than $\lambda_L$ \cite{Sakai_1993,Kleiner_1994}. 

Planar JJs have a coplanar capacitor geometry. In this case the electric field is not locked in the junction and decays slowly (quadratically) along the electrode \cite{Murray_2018}, as shown by red lines in Fig. \ref{fig:fig5} (b). Magnetic field is applied perpendicular to electrodes \cite{Boris_2013}. In this case the demagnetization factor is huge, resulting in appearance of profound stray fields \cite{Golod_2019,Golod_2019B}. The Meissner effect in superconducting electrodes does not allow closing of stray fields through the electrodes. To close, they have to stretch all the way to the neighbor junctions, as shown by green lines in Fig. \ref{fig:fig5} (b), no mater how long is the separation \cite{Golod_2019B}. 
Thus, unlike overlap JJs, planar JJs can interact directly via stray electric and magnetic fields. The range of such interaction is not related to any screening length and is determined solely by the geometry of the array.    
Moreover, the long-range expansion of stray fields into open space and substrate allows effective interaction of planar JJs with external circuits and, thus, amplify both electromagnetic wave emission \cite{Boris_2013} and indirect coupling via external cavities \cite{Almaas_2002}. The discussed difference between overlap and planar JJs is, essentially, a consequence of local and non-local electrodynamics in the respective JJs \cite{Mintz_2001,Boris_2013}. 

\section{Conclusions}
To conclude, we fabricated and studied experimentally arrays of planar Josephson junctions with access to intermediate electrodes. This allowed detailed insight into intricate internal
dynamics of the arrays. We observed a surprisingly strong mutual
coupling, leading to a robust phase-locking, despite a substantial separation between junctions. 
Dynamic (via induced ac-currents) and static (via stray flux of Josephson vortices) coupling mechanisms were identified. We argued that the direct long-range interaction between planar junctions is mediated  by non-local and long-range stray electromagnetic fields. This facilitates synchronization of large arrays required for coherent operation of multi-junction electronics. Such coherent electronics may provide significant improvements and advantages. For example, synchronization of JJs allows multiplication enhancement of the read-out voltage, proportional to the number of phase-locked junctions. We have demonstrated the $I_c R_n$ multiplication in excess of 6 mV for just 9 JJs. Furthermore, long-range extension of stay electromagnetic fields in planar junctions helps both in synchronization of many junctions and in propagation of the electromagnetic power into open space. This is beneficial for operation of coherent Josephson oscillators. 
We conclude that the planar geometry is advantageous for realization of various types of coherent Josephson electronics. 

\begin{acknowledgments}
The work was supported by the Russian Science Foundation Grant No.
19-19-00594 (sample fabrication and experiment) and 20-42-04415 (data analysis and manuscript preparation). 
\end{acknowledgments}


%


\begin{thebibliography}{99}


\bibitem{Jain_1984} A. K. Jain, K. K. Likharev, J.E. Lukens, and J,E, Sauvageau, Mutual phase-Locking in Josephson junction arrays. {\em Phys. Rep. } {\bf 109}, 309-426 (1984).

\bibitem{Han_1994} S. Han, B. Bi, W. Zhang, and J. E. Lukens, Demonstration of Josephson effect submillimeter wave sources with increased power.  {\em Appl. Phys. Lett. } {\bf 64}, 1424 (1994).

\bibitem{Barbara_1999} P. Barbara, A.B. Cawthorne, S.V. Shitov, \&
C.J. Lobb, Stimulated emission and amplification in Josephson
junction arrays. {\em Phys. Rev. Lett.} {\bf 82}, 1963 (1999).

\bibitem{Kleiner_2000} Kleiner, R., Gaber, T. \& Hechtfischer, G. Stacked long Josephson junctions in zero magnetic field: A numerical study of coupled one-dimensional sine-Gordon equations. {\em Phys. Rev. B} {\bf 62}, 4086 (2000).

\bibitem{Filatrella_2006} Filatrella, G., Pedersen, N. F. \& Wiesenfeld, K. Synchronization of Josephson vortices in multi-junction systems. {\em Physica C} {\bf 437-438}, 65 (2006).

\bibitem{Ozyuzer_2007} L. Ozyuzer, A. E. Koshelev, C. Kurter, N. Gopalsami, Q. Li, M. Tachiki, K. Kadowaki,
T. Yamamoto, H. Minami, H. Yamaguchi, T. Tachiki, K. E. Gray,
W.-K. Kwok, and U. Welp, Emission of Coherent THz Radiation from
Superconductors. {\em Science} {\bf 318}, 1291 (2007).



\bibitem{Benseman_2013} T. M. Benseman, K. E. Gray, A. E. Koshelev, W.-K. Kwok, U. Welp, H. Minami, K. Kadowaki, and T. Yamamoto, Powerful terahertz emission from Bi$_2$Sr$_2$CaCu$_2$O$_{8+\delta}$ mesa arrays. {\em Appl. Phys. Lett.} {\bf 103}, 022602 (2013).

\bibitem{Welp_2013} U. Welp, K. Kadowaki, and Kleiner, Superconducting emitters of THz radiation.
{\em Nature Photonics} {\bf 7}, 702 (2013).

\bibitem{Borodianskyi_2017} E. A. Borodianskyi and V. M. Krasnov, Josephson emission with
frequency span 1-11 THz from small
Bi$_2$Sr$_2$CaCu$_2$O$_{8+\delta}$ mesa structures, {\em Nat.
Commun.} {\bf 8}, 1742 (2017).

\bibitem{Galin_2018} M. A. Galin, E. A. Borodianskyi, V. V. Kurin, I. A. Shereshevskiy, N. K. Vdovicheva,
V. M. Krasnov, and A. M. Klushin, Synchronization of Large
Josephson-Junction Arrays by Traveling Electromagnetic Waves. {\em
Phys. Rev. Appl.} {\bf 9}, 054032 (2018).

\bibitem{Cattaneo_2021} R. Cattaneo, E.A. Borodianskyi, A.A. Kalenyuk, V. M. Krasnov, Superconducting THz sources with 12$\%$ power efficiency. {\em
Phys. Rev. Appl.} {\bf 16}, L061001 (2021).


\bibitem{Mros_1998} N.Mros, V. M. Krasnov, A.Yurgens, D. Winkler and T. Claeson, Multiple-valued c-axis critical current and phase locking in Bi$_2$Sr$_2$CaCu$_2$O$_{8+\delta}$ single crystals. {\em Phys. Rev. B} {\bf 57}, R8135 (1998).




\bibitem{Klushin_1995} A. M. Klushin and H. Kohlstedt, Experimental study on stacked Josephson tunnel junction arrays under microwave irradiation, {\em J. Appl. Phys.} {\bf 77}, 441 (1995).

\bibitem{Ravindran_1996} K. Ravindran, L. B. Gomez, R. R. Li, S. T. Herbert, P. Lukens, Y. Jun, S. Elhamri, R. S. Newrock, and D. B. Mast, Frequency dependence of giant Shapiro steps in ordered and site-disordered proximity-coupled Josephson-junction arrays, {\em Phys. Rev. B } {\bf 53}, 5141 (1996).


\bibitem{Golod_2019} T. Golod, O. M. Kapran, and V. M. Krasnov, Planar Superconductor-Ferromagnet-Superconductor Josephson Junctions as Scanning-Probe Sensors. {\em Phys. Rev. Appl.} {\bf 11}, 014062
(2019).

\bibitem{Cybart_2019} J. C. LeFebvre, E. Cho, H. Li, K. Pratt, and S. A. Cybart, Series arrays of planar long Josephson junctions for high dynamic range magnetic flux detection. {\em AIP Advances} {\bf 9}, 105215 (2019).

\bibitem{Krasnov_2002} V. M. Krasnov, Stacked Josephson junction SQUID. {\em Physica C} {\bf 368}, 246-250 (1992).


\bibitem{Sakai_1993} S. Sakai, P. Bodin, and N.F. Pedersen, Fluxons in thin-film superconductor-insulator superlattices. {\em J. Appl. Phys.} {\bf 73}, 2411 (1993).

\bibitem{Kleiner_1994} R. Kleiner, Two-dimensional resonant modes in stacked Josephson junctions. {\em Phys. Rev. B} {\bf 50}, 6919 (1994).


\bibitem{Koyama_1996} T. Koyama and M. Tachiki, $I-V$ characteristics of Josephson-coupled layered superconductors with longitudinal plasma excitations. {\em Phys. Rev. B} {\bf 54}, 16183 (1996).

\bibitem{Shukrinov_2007} Yu.M. Shukrinov and F. Mahfouzi, Influence of Coupling between Junctions on Breakpoint Current in Intrinsic Josephson Junctions. {\em Phys. Rev. Lett.} {\bf 98}, 157001 (2007).



\bibitem{Ustinov_1993} A. V. Ustinov, H. Kohlstedt, M. Cirillo, N. F. Pedersen, G. Hallmanns, Coupled fluxon modes in stacked Nb/AlO$_x$ /Nb long Josephson junctions.  {\em Phys. Rev. B} {\bf 48}, 10614 (1993).

\bibitem{Nevirkovets_1994} I.P. Nevirkovets, J.E. Evetts, M.G. Blamire, Transition from single junction to double junction behaviour in SISIS-type Nb-based devices.  {\em Phys. Lett. A} {\bf 187}, 119-126 (1994).


\bibitem{Almaas_2002} E. Almaas and D. Stroud, Dynamics of a Josephson array in a resonant cavity. {\em Phys. Rev. B } {\bf 67}, 134502 (2002).

\bibitem{Galin_2020} M. A. Galin, F. Rudau, E. A. Borodianskyi, V.V. Kurin, D. Koelle, R. Kleiner,
V.M. Krasnov, and A.M. Klushin, Direct Visualization of Phase-Locking of Large Josephson Junction Arrays by Surface Electromagnetic Waves. {\em Phys. Rev. Appl.} {\bf 14}, 024051
(2020).


\bibitem{Kohlstedt_1995} H. Kohlstedt, A. V. Ustinov,  and F. Peter, Double Barrier Long Josephson Junctions with a Contact to the Intermediate Superconducting Layer.  {\em IEEE Trans. Appl. Supercond. } {\bf 5}, 2939 (1995).

\bibitem{Darula_1995} M. Darula, S. Beuven, M. Siegel, A. Darulova and P. Seidel, Phase locking in a multijunction superconducting loop. {\em Appl. Phys. Lett. } {\bf 67}, 1618 (1995).


\bibitem{Goldobin_1996} E. Goldobin, H. Kohlstedt, and A. V. Ustinov, Tunable phase locking of stacked Josephson flux-flow oscillators.  {\em Appl. Phys. Lett.} {\bf 68}, 250 (1996).

\bibitem{Mintz_2001} V. G. Kogan, V.V. Dobrovitski, J. R. Clem, Y. Mawatari,
and R. G. Mints, Josephson junction in a thin film. {\em Phys. Rev. B} {\bf 63}, 144501 (2001).

\bibitem{Boris_2013} A. A. Boris, A. Rydh, T. Golod, H. Motzkau, A. M. Klushin,
and V. M. Krasnov, Evidence for Nonlocal Electrodynamics
in Planar Josephson Junctions. {\em Phys. Rev. Lett.} {\bf 111}, 117002
(2013).

\bibitem{Golod_2019B} T. Golod, A. Pagliero, and V. M. Krasnov, Two mechanisms of
Josephson phase shift generation by an Abrikosov vortex. {\em Phys.
Rev. B} {\bf 100}, 174511 (2019).

\bibitem{Krasnov_2005} V.M. Krasnov, O. Ericsson, S. Intiso, P. Delsing, V.A. Oboznov,
A.S. Prokofiev, V.V. Ryazanov, Planar S–F–S Josephson junctions made by focused ion
beam etching. {\em Physica C} {\bf 418}, 16-22 (2005).

\bibitem{Octavio_1988} K. Flensberg, J. Bindslev Hansen, and M. Octavio, Subharmonic energy-gap structure in superconducting weak link. {\em Phys. Rev. B} {\bf 38}, 8707 (1988).

\bibitem{Kapran_2021} O. M. Kapran, R. Morari, T. Golod, E. A. Borodianskyi,
V. Boian, A. Prepelita, N. Klenov, A. S. Sidorenko and V. M. Krasnov, In situ transport characterization of magnetic states in Nb/Co
superconductor/ferromagnet heterostructures. {\em Beilstein J. Nanotechnol.} {\bf 12}, 913–923 (2021).

\bibitem{Supplem} See the supplementary material at ... . The supplementary contains two video files, which show details of magnetic field variation of the $I$-$V$s for the two arrays, as in Figs. \ref{fig:fig2} and \ref{fig:fig3}.

\bibitem{Barone} A. Barone and C. Paterno, Physics and Applications of the
Josephson Effect. (J. Wiley $\&$ Sons, New York, USA, 1982).


\bibitem{Kulik_1978} I. Kulik, A. Omelyanchouk. The Josephson effect in superconducting constrictions: Microscopic theory. {\em J. de Physique Colloques} {\bf 39 (C6)}, C6-546-C6-547 (1978). 


\bibitem{Zdravkov_2010} V. I. Zdravkov, J. Kehrle, G. Obermeier, S. Gsell, M. Schreck, C. M\"{u}ller, H.-A. Krug von Nidda, J. Lindner, J. Moosburger-Will, E. Nold, R. Morari, V. V. Ryazanov, A. S. Sidorenko, S. Horn, R. Tidecks, and L. R. Tagirov, Reentrant superconductivity in superconductor/ferromagnetic-alloy bilayers.  {\em Phys. Rev. B} {\bf 82}, 054517 (2010).

\bibitem{Blamire_2014} N. Banerjee, J. W. A. Robinson, and M. G. Blamire, Reversible
control of spin-polarized supercurrents in ferromagnetic
Josephson junctions, {\em Nat. Commun.} {\bf 5}, 4771 (2014).


\bibitem{Birge_2016} E. C. Gingrich, B. M. Niedzielski, J. A. Glick, Y. Wang, D. L. Miller, R.Loloee,
W. P. Pratt Jr. and N. O. Birge, Controllable $0–\pi$ Josephson junctions containing a
ferromagnetic spin valve. {\em Nature Physics} {\bf 12}, 564 (2016).



\bibitem{Dremov_2019} V. V. Dremov, S. Y. Grebenchuk, A. G. Shishkin, D. S. Baranov, R. A. Hovhannisyan, O. V. Skryabina, I. A. Golovchanskiy, V. I. Chichkov, C. Brun, T. Cren, V. M. Krasnov, A. A. Golubov, D. Roditchev, and V. S. Stolyarov, Local Josephson vortex generation and manipulation with a magnetic force microscope. {\em Nat. Commun.} {\bf 10}, 4009 (2019).


\bibitem{Murray_2018} C. E. Murray, J. M. Gambetta, D. T. McClure, and M. Steffen, Analytical Determination of Participation in Superconducting Coplanar Architectures. {\em IEEE Trans. Microwave Theory and Techniques} {\bf 66}, 3724 - 3733 (2018).  










\end{thebibliography}
\end{document}